\documentclass[aps,prl,twocolumn,amsmath,
amssymb,10pt]{revtex4-1}
\usepackage{graphicx}
\usepackage{ulem}
\usepackage{soul}
\usepackage{txfonts}
\usepackage[colorlinks, linkcolor=blue]{hyperref}
\usepackage{verbatim}
\usepackage{esint}
\usepackage{amsmath}
\usepackage{mathtools}
\renewcommand{\vec}[1]{\boldsymbol{#1}}

\def \a{{\vec a}}

\def \e {{\vec e}}

\def \ve {\varepsilon}
\def \r {{\vec r}}

\def \s {\psi}
\def \q {{\vec q}}
\def \d{\partial}

\def \ve {\varepsilon}

\def \ac {{\cal{S}}}
\def \z {\zeta}

\def \D{\Delta}
\def \B{B_\perp}
\def \A{{\bf A}}

\def \L{{\cal{L}}}

\def \beq {\begin{eqnarray}}
\def \eeq {\end{eqnarray}}
\def \tn {\textnormal}

\def \M {\vec {M}}

\def \M {M\overline{M}}

\def \ua{\uparrow}
\def \da{\downarrow}
\def \lb{\ell_B}

\begin{document}

\title{Effect of magnetization on the tunneling anomaly in compressible quantum Hall states}
\author{Debanjan Chowdhury}
\thanks{These two authors contributed equally}
\author{Brian Skinner}
\thanks{These two authors contributed equally}
\author{Patrick A. Lee}
\affiliation{Department of Physics, Massachusetts Institute of Technology, Cambridge Massachusetts
02139, USA.}

\date{\today \\
\vspace{.1in}}

\begin{abstract}

Tunneling of electrons into a two-dimensional electron system is known to exhibit an anomaly at low bias, in which the tunneling conductance vanishes due to a many-body interaction effect. Recent experiments have measured this anomaly between two copies of the half-filled Landau level as a function of in-plane magnetic field, and they suggest that increasing spin polarization drives a deeper suppression of tunneling.
Here we present a theory of the tunneling anomaly between two copies of the partially spin-polarized Halperin-Lee-Read state, and we show that the conventional description of the tunneling anomaly, based on the Coulomb self-energy of the injected charge packet, is inconsistent with the experimental observation.  
We propose that the experiment is operating in a different regime, not previously considered, in which the charge-spreading action is determined by the {\it compressibility} of the composite fermions.    
\end{abstract}

\maketitle

{\it Introduction.-} The tunneling of electrons into a metal is known to exhibit a  ``tunneling anomaly" (TA), in which electron-electron interactions cause the tunneling conductance to vanish continuously as the bias voltage is brought to zero. Conceptually, the tunneling process can be separated into two distinct steps: (1) a fast, `single-particle' transmission of an electron across the tunneling barrier, and (2) a slower, `many-body' process in which the electronic fluid in the metal rearranges to accommodate the extra electron [as depicted in Fig. \ref{expt}(a)]. At low voltages the latter process acts as a bottleneck, and therefore effectively determines the tunneling rate and the tunneling conductivity.  For this reason a measurement of the TA can be used to probe the nature of interactions in an electron system.

In the half-filled Landau level of a two-dimensional electron system, electrons realize a particularly interesting and strongly-correlated metallic phase. The lack of a quantized Hall effect at filling factor $\nu=1/2$ can be understood within the framework of composite fermions (CFs) \cite{JKJ}, where each electron is attached to two flux quanta.  This state was described by Halperin, Lee, and Read (HLR) \cite{HLR} in terms of  a low energy effective field theory for the CFs coupled to an emergent gauge field with a Chern-Simons (CS) term. At filling factor $\nu=1/2$, the CFs see no magnetic field on average and form a Fermi surface. When an electron tunnels into the half-filled Landau level, it is this CF fluid whose many-body rearrangement provides the bottleneck for tunneling.  One can therefore expect that the tunneling conductance into the $\nu = 1/2$ state is influenced by a combination of the state's properties, including the charge conductivity, the electron-electron interaction strength, and the compressibility.

The tunneling between two quantum Hall systems with total filling factor $\nu_T=1$  has attracted particular interest during the past three decades, with experiments showing clear evidence for a TA \cite{Ashoori, Eis92, Jones94}. Theoretical explanations for this anomaly have focused primarily on the limit of spatially well-separated layers, and have assumed complete spin polarization \cite{HPH, XGW94, LS}.  Numerous studies during the past two decades, however, have shown that at low electron density the half-filled Landau level is not fully spin polarized \cite{Kukushkin, Dementyev, Shayegan, Spielman, Kumada, Dujovne, Tracy,  Giudici, Smet1, Smet2, Finck} . 
A very recent experiment \cite{JPE16} has returned to the problem of the TA in bilayers with total filling $\nu_T=1$, focusing on the role of spin polarization in bilayers with relatively small spacing $d$.  The authors of \cite{JPE16} found that, as the spin polarization is increased using an in-plane magnetic field, the tunneling conductance is increasingly suppressed [Fig.\ \ref{expt}(b)]. 

\begin{figure}
\begin{center}
\includegraphics[width=1.0\columnwidth]{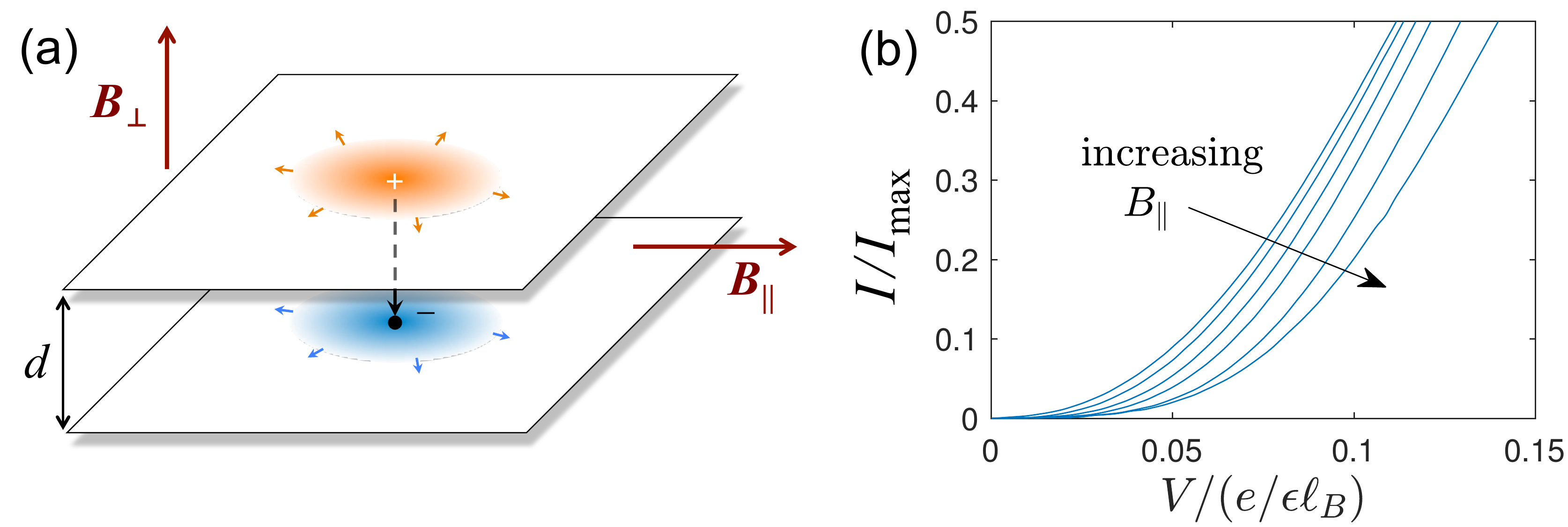}
\includegraphics[width=1.0\columnwidth]{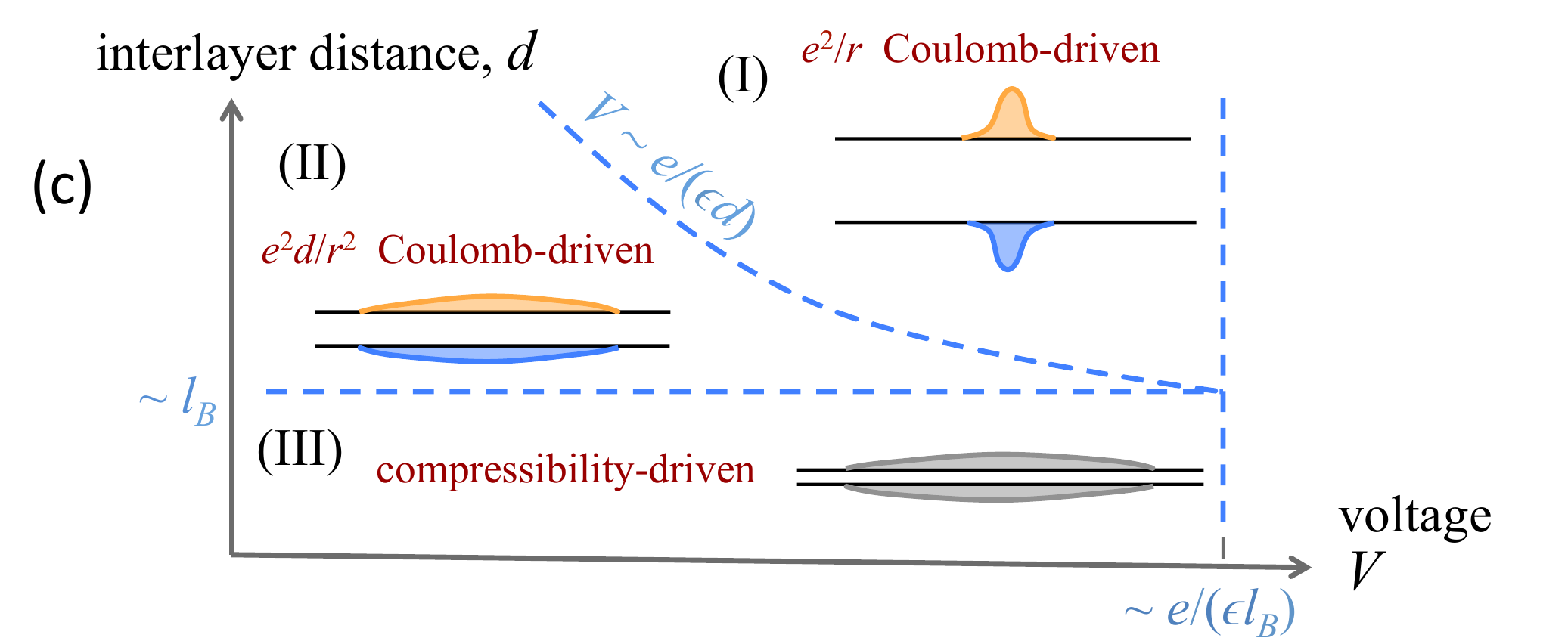}
\end{center}
\caption{(a) Schematic illustration of the tunneling process.  At low bias voltage, the tunneling of an electron from the top layer to the bottom layer must be accompanied by an outward spreading of positive (negative) charge in the top (bottom) layer. (b) Experimentally-measured tunneling current \cite{JPE16} as a function of voltage for $d = 1.96 \lb$, where $\lb = \sqrt{\hbar c/eB_\perp}$.  From left to right, different lines correspond to $B_{\parallel}$ ranging from $0$ to $3.23$\,T, with $B_{\perp} = 3.23$\,T held constant.  (c) Schematic map of the three regimes (labeled $\rm{I}, \rm{II}, \rm{III}$) described in this work.  Inset images depict, schematically, the scale of the spreading charge relative to the layer separation.}
\label{expt}
\end{figure}

In this paper we focus on the TA in quantum hall bilayers at low bias voltage, and we show that the suppression of tunneling with increasing spin polarization is inconsistent with previous theoretical treatments of the TA, which predict an \textit{increase} in tunneling current with spin polarization. Thus, an explanation of the experimental data apparently requires us to consider a qualitatively new regime. We compute the one-electron spectral function that describes the tunneling of electrons in a quantum Hall bilayer at $\nu_T=1$, and we show that its behavior can be understood in terms of three regimes [summarized graphically in Fig.\ \ref{expt}(c)].

These regimes can be understood qualitatively as follows. The many-particle rearrangement that accompanies electron tunneling is characterized by a typical length scale $r$, which describes the spatial extent of the perturbation of charge density in the two layers, and a typical energy $U(r) \sim e V$, where $V$ is the bias voltage.  At large inter-layer spacing $d$ (regime I), the Coulomb-energy is dominated by the intra-layer Coulomb interaction, and $U_\textrm{I}(r) \sim e^2/(\epsilon r)$, where $-e$ is the electron charge and $\epsilon$ is the dielectric constant.  When $d$ is reduced to the point that $r \gg d$, inter-layer interactions become important (regime II), and the Coulomb energy of the two spreading charge packets becomes similar to that of a plane capacitor: $U_\textrm{II}(r) \sim e^2 d/(\epsilon r^2)$.  Equating $U_\textrm{I}(r)$ and $U_\textrm{II}(r)$, and using $U(r) \sim eV$, implies that the boundary between these two regimes is described by $V \sim e/(\epsilon d)$. As we show below, neither regime I nor II is consistent with the experiments of Ref.\ \cite{JPE16}. However, if $d$ is made very small (regime III), then the Coulomb energy of the spreading charge is quenched, and $U(r)$ is instead dominated by the energy associated with the finite compressibility of the spreading charge packet, $U_\textrm{III}(r) \sim \xi_d/r^2$. Here $\xi_d \sim \hbar^2/m^*$ is the compressibility, with $m^*$ the effective mass of the CFs. Since $\hbar^2/m^*$ is of order $e^2 \lb/\epsilon$ in the HLR state, the boundary between regimes II and III corresponds to $d/\lb$ reaching a constant of order unity.

We note that our focus is on low voltages, $V \ll e/(\epsilon \lb)$, where the current is far below its peak value $I_\tn{max}$.  The behavior of the peak current was considered in Ref.\ \cite{JKJ17}, where the evolution of the peak with in-plane magnetic field was explained in terms of the field-dependent shift in the position of the guiding center of the tunneled electron. This shift is not relevant for the TA, since at low voltage the length scale $r$ is much longer than the magnitude of the shift.

{\it Model.-} 
Let $\s_{e,s,\sigma}(\r)$ and $\s_{s,\sigma}(\r)$ represent the electron and CF annihilation operators, respectively, at position $\r$ in layer $s(= 1,2)$, with spin quantum number $\sigma(=\ua,\da)$. Let $\rho_{s,\sigma}(\r)$ be the density of electrons (or, equivalently, CFs) with spin $\sigma$ in layer $s$. We attach flux to the electrons such that a CF of any spin orientation sees $\phi$ flux quanta attached to electrons of both spin components in the {\it same} layer and no flux quanta attached to electrons in the {\it opposite} layer \cite{NB93}. The global densities of electrons in a given layer, $n_{\ua(\da)}$, are such that $n_\ua + n_\da = n$ and $n_\ua - n_\da = \zeta n$, where $n$ is the total electron concentration and $\zeta$ is the relative polarization.  

Each CF then sees an effective average field $\overline{\B} = \B -2\pi\phi n/e$. We are interested in the problem where each layer is at $\nu=\nu_\ua + \nu_\da = nhc/e\B=1/2$, i.e. where $\B=4\pi n/e$; the unique choice for doing this is when $\phi=2$. The CFs (of either spin component) do not see a magnetic field on average and they form Fermi surfaces in each layer with Fermi wave vectors $k_{F\uparrow (\downarrow)} = \sqrt{4\pi n_{\uparrow(\downarrow)}}$ \cite{JKJspin}. We note that our results below can be generalized in a straightforward fashion to other even-denominator gapless, spin-polarized filling fractions. All of the regimes described above remain qualitatively similar but the numerical prefactors of the tunneling exponents will be different.

The low-energy field theory for the CF Fermi surfaces minimally coupled to the gauge field is given by \cite{HLR,NB93}
\beq
\label{CFlag}
\L &=& \L_0 + \L_{\tn{int}} + \L_{\tn{CS}},\\
\L_0 &=& \sum_{s,\sigma} \bigg(\s_{s,\sigma}^\dagger(\r,\tau)[\d_\tau + i a_0^s(\r,\tau)]\s_{s,\sigma}(\r,\tau) \nonumber\\
&+& \frac{1}{2m^*_\sigma}\s_{s,\sigma}^\dagger(\r,\tau) [-i\nabla+\D \a^s(\r,\tau)]^2 \psi_{s,\sigma}(\r,t)  \bigg),\nonumber\\
\L_{\tn{int}} &=& \sum_{s,s'} \frac{1}{2}\int_\r \int_{\r'} V_{s,s'}(\r-\r') :\rho_s(\r) \rho_{s'}(\r'): 
\nonumber
\eeq
where $m^*_{\uparrow(\downarrow)}$ denote the effective masses for the different spin-components, $\D\a$ denotes the gauge field minus $e \A$, with $\A$ being the external vector potential, and `$:~:$' denotes normal ordering. The Coulomb interaction, $V_{s,s'}(\r) = 2\pi e^2/\left(\epsilon \sqrt{r^2 + d^2(1-\delta_{s,s'})}\right)$, is insensitive to the spin label. 

The Chern-Simons term is
\beq
\L_{\tn{CS}} &=& -\frac{i}{2\pi}\sum_{ss'} \int_{\r} K^{-1}_{ss'} ~a_0^s(\r,\tau)~ \hat{z}\cdot[\nabla\times \a^{s'}(\r,\tau)],
\label{CS}
\eeq
where, as discussed earlier, $K_{ss'}$ is diagonal with respect to the layer-index: $K_{ss'} = \phi~ \delta_{ss'}$. Integrating out $a_0^s(\r,\tau)$ from the action leads to the constraint
\beq
\sum_\sigma\rho_{s,\sigma}(\r,t)  =  \frac{\hat{z}\cdot \nabla\times \a^s(\r,\tau)}{2\pi\phi} 
\equiv \frac{b^s(\r,\tau)}{2\pi\phi}.
\label{constraint}
\eeq
That is, $\phi$ fictitious $\a^s$ flux quanta are attached to both spin species in each layer and $b^s = (\d_x a_y^s - \d_y a_x^s)$ is the magnetic field associated with the internal gauge field.

{\it Spectral function.-} The single-electron Green's function associated with tunneling an electron with spin $\sigma$ into layer $s$ at $\r=0$ and time $t=0$ and then removing an electron at $\r=0$ with the same spin and from the same layer at a later time $t=\tau$ is given by $G_{s,\sigma}(\tau) =\langle \psi_{e,s,\sigma}(\vec{0},\tau)~\psi_{e,s,\sigma}^\dagger(\vec{0},0)\rangle$,
\beq
G_{s,\sigma}(\tau) =  \int {\cal{D}}[\psi~a]~ \psi_{s,\sigma}(\tau)~\psi_{s,\sigma}^\dagger(0)~\delta(\M)\nonumber\\
 ~\tn{exp}(-S[\psi^\dagger,\psi,a_\mu]),
\label{Gst}
\eeq
where $S[\psi^\dagger,\psi,a_\mu]$ is the imaginary-time action corresponding to the field theory introduced in Eq.\ (\ref{CFlag}). 
Here, $\delta(\M)$ denotes the boundary condition in space-time on the gauge field, corresponding to creating and annihilating two flux quanta, and the path integral measure ${\cal{D}}[\psi~a] \equiv \prod_{s',\sigma'}~D\psi_{s',\sigma'}^\dagger~D\psi_{s',\sigma'}~Da^{s'}_\mu$. In the path integral, the above boundary condition can be equivalently interpreted \cite{XGW94} as inserting and subsequently removing a doubly-charged monopole
\footnote{We note that unlike the purely (2+1)-dimensional compact U(1) gauge theory, which is unstable to confinement at the longest length scales \cite{polyakov}, the action described above has a deconfined phase with well defined CF excitations.}.

The Green's function in Eq.\ (\ref{Gst}) can be re-expressed as a path integral over the CF fields with the configuration of $a_\mu$ held fixed and a path integral over all allowed configurations of $a_\mu$ subject to the appropriate boundary conditions. 
For the bilayer problem, the boundary condition requires a current that sources the internal gauge field,
\beq
j_\mu^1 &=& [\theta(x_0-\tau) - \theta(x_0)] ~\delta^{(2)}(\r) ~\delta_{\mu0} 
\label{BP2}
\eeq
for the top layer and $j_\mu^2 = -j_\mu^1$ for the bottom layer, which corresponds to the creation of a monopole in the top and an anti-monopole in the bottom layer at time $t=0$, both of which are removed at a later time $\tau$ at the same position $\r=0$. In the limit of times much longer than the inverse Fermi energy, this process couples only to the low-energy diffusive mode \cite{HLR,NB93} with $\omega\sim V(q) q^3$, where $V(q) = 2\pi e^2(1-e^{-qd})/(\epsilon q)$. 

Interestingly, the boundary condition of Eq.\ (\ref{BP2}) does not contain any information about the spin of the injected electron; the inserted monopole/antimonopole does not have a spin quantum number. The constraint associated with the flux attachment [Eq.\ (\ref{constraint})] dictates the total CF density but gives no information about the magnetization of the perturbation. In general, this magnetization (the spin composition of the spreading charge) is not simply equal to that of the injected electron, as one might naively expect.  This is because the CS field couples the spin-up and spin-down currents to each other, such that a CF current of either spin gives rise to a transverse CS gauge field that is felt by both spin components.  In this way any perturbation of CF density, regardless of its initial spin composition, quickly evolves to contain a mixture of both components that may not reflect the magnetization $\z$ of the background.

In the limit where the charge spreading is driven purely by the Coulomb energy of the perturbation, the magnetization of the perturbation is irrelevant for the charge spreading, since the Coulomb interaction is independent of spin. However, this is not the case in the regime where the dominant energy scale driving the charge spreading is provided by the finite compressibility of the CF fluid. Instead, in the long-time limit the magnetization of the perturbation is determined by the ratio of the different spin compressibilities, as we show below.

In order to incorporate the dynamic evolution of the spin degree of freedom and the associated magnetization, we introduce the field $\delta m(\r,\tau)$ subject to the following constraint
\beq
\delta\rho_\ua(\r,\tau) - \delta\rho_\da(\r,\tau) &=& \delta m(\r,\tau),\\
\delta\rho_\ua(\r,\tau) + \delta\rho_\da(\r,\tau) &=& \delta n(\r,\tau) = \frac{\delta b(\r,\tau)}{2\pi\phi},
\label{consm}
\eeq
where $\delta\rho_{\ua(\da)} = \rho_{\ua(\da)} - n_{\ua(\da)}$, so that $\delta b/(2\pi\phi)$ and $\delta m$ represent the deviation of the density and the magnetization, respectively, from the homogeneous ground state. Introducing the field $\delta m$ implies an additional contribution to the action $S_\tn{M}[\delta m]$, which we leave unspecified for the time being. The Green's function is then given by
\beq
&&G_{s,\sigma}(\tau)= \int {\cal{D}}[a]~{\cal{D}}[\delta m]~\delta(\M) ~\langle  \psi_{s,\sigma}(\tau)\psi_{s,\sigma}^\dagger(0)\rangle_a \nonumber\\
&&~~~~~~~~~~~~~\tn{exp}(-S_{\tn{eff}}[a_\mu]-S_{\tn{M}}[\delta m]),\\
&&\tn{exp}(-S_{\tn{eff}}[a_\mu]) \equiv  \int {\cal{D}}[\psi]~\tn{exp}(-S[\psi^\dagger,\psi,a_\mu]).
\label{seff}
\eeq
We now assume that the low-energy suppression of the spectral function arises predominantly from the exponential saddle point contribution, $S_{\tn{eff}}[a_\mu,j_\mu,\delta m] = S_{\tn{eff}}[a_\mu] + S_\tn{M}[\delta\overline{m}] - \int_\r a^s_\mu j^s_\mu$ evaluated at the value of $a_\mu = \overline{a}_\mu$ that incorporates the boundary condition (see Supplementary Material for details).  That is, $G_{s,\sigma}(\tau) \approx \tn{exp}(-S_{\tn{eff}}[\overline{a}_\mu,j_\mu,\delta m])~G_0(\tau)$, where $G_0(\tau)$ can be at most an algebraically decaying function of $\tau$ .  

To obtain $S_\tn{eff}[a_\mu]$, we integrate out the CFs and obtain within RPA the effective action \cite{HLR} of the form $S_\tn{eff}[a] = S_\tn{em} + S_{\tn{CS}}$, where
\beq
S_\tn{em} = \frac{1}{2}\sum_{i\omega_n}\int_\q \bigg[ \ve(\q,\omega) |\e_{\q,\omega}|^2 + \beta(\q,\omega) |b_{\q,\omega}|^2\bigg],
\label{effact}
\eeq 
where $i\omega_n$ are the Bosonic Matsubara frequencies and $e_\alpha = \d_0 a_\alpha - \d_\alpha a_0$ is the electric field associated with the internal gauge field. The effective dielectric function, $\ve(\q,\omega)$, and inverse magnetic permeability, $\beta(\q,\omega)$, derive their momentum and frequency dependence from the underlying CF Fermi surfaces:
\beq
\ve(\q,\omega) &=& \frac{2(k_{F\uparrow} + k_{F\downarrow})}{4\pi |\omega_n| q} = \frac{2k_F}{4\pi |\omega_n| q}~g(\zeta),\\
\beta(\q,\omega) &=& \chi_d + \frac{1}{(2\pi\phi)^2} V(\q),\eeq
where $g(\zeta) = \sqrt{(1+\zeta)/2} + \sqrt{(1-\zeta)/2}$ and $\chi_d = (\partial\mu/\partial n)/(2\pi\phi)^2$, where $\mu$ is the chemical potential. {\footnote{In the RPA treatment of HLR theory (for $\zeta=1$ and $m^*_{\uparrow} \equiv m^*$), $\chi_d = 1/(12\pi m^*)$ \cite{HLR}. However, there can be additional corrections to $\chi_d$ as a result of interactions, as we discuss below.}} 

Following Refs.\ \cite{Tru93,XGW94}, and for the boundary condition in Eq.\ (\ref{BP2}), the action is given by
\beq
S_{\tn{eff}}(\tau) = (2\pi\phi)^2~\int_\omega \int_\q \frac{\ve(\q,\omega) ~\beta(\q,\omega)}{\beta(\q,\omega)~ q^2 + \ve(\q,\omega)~\omega^2} (1-\cos(\omega\tau)).\nonumber\\
\label{actMCS}
\eeq
Finally, the total action of the system is obtained by subtracting the action associated with the work performed by the voltage source from the action computed above,
\beq
\ac(\tau) = S_{\tn{eff}}(\tau) - eV\tau.
\eeq
Optimizing the above action over $\tau$ gives an optimal time $\tau_*(V)$ that characterizes the charge accommodation time, and the tunneling conductivity is given by $\sim \exp[-\ac(\tau_*(V))/\hbar]$. We have arrived at the same results for the tunneling action using a complementary, semi-classical hydrodynamic description for the spreading charge \cite{LS} in an accompanying paper \cite{DCBSPL2}.

We now consider the various parametric regimes for the tunneling action.

{\it Large layer separation.-} Let us first consider the regime of large layer separation (region I), where $qd\gg1$, with $q^{-1} \sim r$.  In this limit $V(\q)\approx 2\pi e^2/\epsilon q$ is singular at small $q$ and $\beta(\q,\omega)\approx V(\q)/(2\pi\phi)^2$. The charge spreading in the two layers decouples and the tunneling action (at zero temperature) is given by
\beq
S(\tau_*(V)) = 2 A~g(\zeta) \frac{e^2/\epsilon l_B}{2eV},
\label{eq:Coulombhigh}
\eeq
where $A=4\pi$ and the extra factor of 2 arises due to the contribution from the two layers. This regime is represented as region-$\rm{I}$ in Fig.\ \ref{expt}(c). Eq.\ (\ref{eq:Coulombhigh}) describes a charge-spreading action that \textit{decreases} with increasing spin polarization $\zeta$.  One can think of this decrease as arising from the increase of the CF conductivity with increasing spin polarization \cite{SimonReview}, which allows the perturbation to spread more quickly and lowers the associated action. This dependence is in the opposite direction as observed in experiment (Fig. \ref{expt}b) \cite{JPE16}.

{\it Small layer separation.-} When the layer separation is small ($qd\ll1$), we can approximate the Coulomb interaction as $V(\q)\approx 2\pi e^2 d/\epsilon$. In this regime, $\beta(\q,\omega)$ is independent of momentum at leading order and we denote it as simply $\beta$. The tunneling action (at zero temperature) then has the form
\beq
S(\tau_*(V)) = 2 C~k_F~g(\zeta)~\sqrt{\frac{\beta}{2eV}},
\label{eq:smd}
\eeq
where $C = (-2^{10}\pi\Gamma^3(-1/3)/3^7)^{1/2}$.
Previous studies \cite{HPH,XGW94} (which assumed $\zeta = 1$) have focused on the regime where the Coulomb energy dominates the compressibility, $V(\q)\gg (2\pi\phi)^2\chi_d$, such that $\beta \approx e^2d/(8\pi\epsilon)$. In this limit (region II), the action takes the form
\beq
S(\tau_*(V)) = 2C~ k_F ~g(\zeta) \sqrt{\frac{e^2 d/8\pi\epsilon}{2eV}}.
\label{eq:Coulomblow}
\eeq
Once again, the action {\it decreases} with increasing $\zeta$ and is at odds with the observations of Ref.\ \cite{JPE16}. The action also increases with increasing $d$.

Let us instead consider the situation where $\chi_d \gg V(\q)/(2\pi\phi)^2$, so that $\beta \simeq \chi_d$. It is important to note that while this regime (region III) corresponds to small $d/\lb$, we are simultaneously assuming that the metallic CF state remains a good description and there is no instability toward excitonic condensation \cite{excrev, EisReview}.  This assumption is equivalent to assuming either that $d/\lb$ remains larger than the critical value associated with exciton instability, or that the temperature is larger than the condensation temperature.

In this limit the two oppositely-charged layers are so close that the Coulomb energy of the perturbation is effectively eliminated, and the action is given by twice of the action for a single (decoupled) layer with no long-range Coulomb repulsion. Of course, there can still be a residual interaction on short length scales between the different spin components of the CFs, which can be described phenomenologically within a Landau Fermi liquid approach with undetermined Landau parameters \cite{PN}. 
We assume rotational invariance and use the dimensionless Landau parameters $F_\ell^{\sigma\sigma'} = \sqrt{m_\sigma^* m_{\sigma'}^*} f_\ell^{\sigma\sigma'}/(2\pi)$
{\footnote{Note that due to a lack of spin-rotation invariance, $F^{\ua\ua}\neq F^{\da\da}$ in general, but $F^{\ua\da} = F^{\da\ua}$.}} .
For our purpose, it is sufficient to consider only the $\ell = 0$ component, corresponding to the compression mode of the Fermi surfaces. Following Landau's expansion to quadratic order, the energy can be written as \cite{SimonReview}

\beq
\delta {\cal{E}}(\delta \rho_\ua,\delta \rho_\da) &=& \pi \frac{(1+F_0^{\ua\ua})}{m_\ua^*} (\delta\rho_\ua)^2 + \pi \frac{(1+F_0^{\da\da})}{m_\da^*} (\delta\rho_\da)^2 \nonumber\\
 &+& 2\pi\frac{F_0^{\ua\da}}{\sqrt{m_\ua^* m_\da^*}} \delta\rho_\ua \delta\rho_\da.
\eeq
Using Eq.\ (\ref{consm}), this can be re-expressed as,
\beq
\delta{\cal{E}}(\delta n,\delta m) &=& \frac{\pi}{4}\bigg[\frac{2}{m_{\tn{eff}}} + \frac{f_0^{\ua\ua} + f_0^{\da\da} + 2f_0^{\ua\da}}{2\pi}\bigg] (\delta n)^2  \nonumber\\
&+& \frac{\pi}{4}\bigg[\frac{2}{m_{\tn{eff}}} + \frac{f_0^{\ua\ua} + f_0^{\da\da} - 2f_0^{\ua\da}}{2\pi}\bigg] (\delta m)^2 \nonumber\\
 &+& \frac{2\pi}{4}\bigg[\frac{1}{m_{\ua}} - \frac{1}{m_{\da}} + \frac{f_0^{\ua\ua} - f_0^{\da\da}}{2\pi}\bigg] \delta n~\delta m ,
\eeq
where we have introduced a {\it reduced} mass, $m_{\tn{eff}} = 2m^*_\uparrow m^*_\downarrow / (m^*_\uparrow + m^*_\downarrow)$ (the factor of $2$ ensures that in the limit of identical masses, $m_{\tn{eff}}=m^*_{\uparrow(\downarrow)})$.

By completing the square for $\delta m$ in the above expansion, one can immediately see that 
\beq
\chi_d =  \frac{1}{32\pi}\bigg[\frac{2}{m_{\tn{eff}}} + \frac{f_0^{s\ua} + f_0^{s\da}}{2\pi} - \frac{\bigg(\frac{1}{m_{\ua}} - \frac{1}{m_{\da}} + \frac{f_0^{s\ua} - f_0^{s\da}}{2\pi}\bigg)^2}{\bigg(\frac{2}{m_{\tn{eff}}} + \frac{f_0^{a}}{2\pi}\bigg)}\bigg],\nonumber\\ 
\eeq
where $f_0^{s\ua(\da)} = f_0^{\ua\ua(\da\da)}+f_0^{\ua\da}$ and $f_0^a = f_0^{\ua\ua} + f_0^{\da\da} - 2f_0^{\ua\da}$. In the limit of complete spin-polarization, $\chi_d$ is determined by the usual compressibility and is proportional to $(1+F_0^{\ua\ua})/m_\ua^*$ \cite{SimonReview}. 

The tunneling action in region III is then given by Eq.\ (\ref{eq:smd}), with $\beta = \chi_d$. In this description, the dependence of the tunneling current on spin polarization $\z$ depends on the way in which the Landau parameters vary with $\z$.  This dependence cannot be known \textit{a priori}, but in principle it can be deduced from experiments, done as a function of $\z$. It is plausible that our description in this regime correctly reproduces the experimental results of Ref.\ \cite{JPE16}, but this remains to be shown experimentally. For example, one can measure the inverse compressibility, which is proportional to $\chi_d$ above, through capacitance \cite{Eis94}.   In addition, it would be interesting to measure the dependence of density on magnetic field at fixed chemical potential that would give a susceptibility inversely proportional to $f_0^{s\ua} - f_0^{s\da}$.

{\it Summary and Outlook.-}
In this paper we have presented a derivation of the action associated with electron tunneling between two compressible CF systems, which determines the tunneling current.  In particular, we have examined the role of incomplete spin polarization across a range of values for the interlayer separation.  One of our main results is that a description where charge spreading is driven primarily by the Coulomb energy of the density perturbation (as in Refs.\ \cite{XGW94, LS, HPH}) is inconsistent with recent experiments \cite{JPE16}. This observation has led us to identify a new regime of behavior for the TA, in which charge spreading is dominated by the finite compressibility of the electron liquid. 

In addition to the experiments we propose above, our results suggest that at small $d/\lb$ the tunneling current should have the functional form $\ln I \propto - 1/\sqrt{V}$ implied by Eq.\ (\ref{eq:smd}). The experimentally measured tunneling current is indeed consistent with this functional form \cite{JPE16,JPE18} at small voltage (see Supplemental Material for details).
Further, the effect of spin polarization on the tunneling current should become weaker with increasing $d/\lb$, as the system moves from the compressibility-dominated regime to the Coulomb-dominated regime.  At large enough $d/\lb \gg 1$ the dependence of tunneling current on spin polarization should reverse sign. Finally, we note that the formalism developed in our paper can be used to describe tunneling experiments in other ``vortex-metals" \cite{VMts}, e.g. in two-dimensional disordered thin film superconductors at large magnetic fields  \cite{kap,SAK}.

{\it Acknowledgments.-} 
We thank J.\ P.\ Eisenstein and I.\ Sodemann for useful discussions. We acknowledge the hospitality of IQIM-Caltech, where a part of this work was completed. DC is supported by a postdoctoral fellowship from the Gordon and Betty Moore Foundation, under the EPiQS initiative, Grant GBMF-4303, at MIT. BS was supported as part of the MIT Center for Excitonics, an Energy Frontier Research Center funded by the U.S. Department of Energy, Office of Science, Basic Energy Sciences under Award no. DE-SC0001088. PAL acknowledges support by DOE under Award no. FG02-03ER46076.

\bibliographystyle{apsrev4-1_custom}
\bibliography{qhe}

\begin{widetext}
\section{Supplementary Material}
\subsection{Computation of Green's function}
We provide here some additional details \cite{XGW94} for the computation of the electronic Green's function in Eq.\ ({\color{blue} 8}). After expressing the Green's function as a path-integral over the CF fields, with a fixed $a_\mu$ background and a path-integral over all allowed configurations of $a_\mu$ subject to the appropriate boundary conditions, it takes the form
\beq
G_{s,\sigma}(\tau)  &=& \int ~{\cal{D}}[a]~{\cal{D}}[\delta m]~\delta(\M) ~\langle  \psi_{s,\sigma}(\tau)~\psi_{s,\sigma}^\dagger(0)\rangle_a~ \tn{exp}(-S_{\tn{eff}}[a_\mu] - S_\tn{M}[\delta m]),~\tn{where}\\
\langle \psi_{s,\sigma}(\tau)~\psi_{s,\sigma}^\dagger(0)\rangle_a &=& \int {\cal{D}}[\psi]~\psi_{s,\sigma}(\tau)~\psi_{s,\sigma}^\dagger(0) ~\frac{\tn{exp}(-S[\psi^\dagger,\psi,a_\mu])}{\tn{exp}(-S_{\tn{eff}}[a_\mu])},
\eeq
where as discussed earlier, we also promote the magnetization $\delta m(\r,\tau)$ to be a dynamical field. The effective action for the gauge-fields takes the form
\beq
\tn{exp}(-S_{\tn{eff}}[a_\mu]) &\equiv&  \int {\cal{D}}[\psi]~\tn{exp}(-S[\psi^\dagger,\psi,a_\mu]).
\eeq
Note that neither of the two terms --- i.e the $\langle \psi \psi^\dagger\rangle$ correlator and the action $\tn{exp}(-S_{\tn{eff}}[a_\mu])$ --- in the expression for $G_{s,\sigma}(\tau)$ above are individually gauge-invariant. However introducing a fermion current source, $j_\mu^s$, that sources the internal gauge-field and modifies the action to $S_\tn{eff}[a_\mu,j_\mu] = S_{\tn{eff}}[a_\mu] - \int_\r a^s_\mu j^s_\mu$ cures this problem. The two terms now in $S_\tn{eff}[a_\mu,j_\mu]$ are not individually gauge-invariant due to the presence of (anti-)monopoles for the first and the current $j_\mu$ not being conserved due to the creation/annihilation of electrons, but the combination is gauge-invariant. 

As discussed in the main text, we focus on the saddle-point contribution, $S_{\tn{eff}}[\overline{a}_\mu,j_\mu,\delta m]$,  around $a_\mu = \overline{a}_\mu$, subject to the $\delta(\M)$ boundary conditions,
\beq
G_{s,\sigma}(\tau) &\approx& \tn{exp}(-S_{\tn{eff}}[\overline{a}_\mu,j_\mu],\delta m)~G_0(\tau),~\tn{where}\\
G_0(\tau) &=& \int {\cal{D}}[\delta a]~\bigg[\langle  \psi_{s,\sigma}(\tau)~\psi^\dagger_{s,\sigma}(0)\rangle_a~\tn{exp}(-\int_\r a^{s'}_\mu j^{s'}_\mu) \bigg]~ \tn{exp}(-S_{\tn{eff}}[\delta a_\mu]),
\eeq
where $\delta a_\mu = a_\mu - \overline{a}_\mu$ and $S_{\tn{eff}}[\delta a_\mu]$ describes the fluctuation of the gauge-field and the Gaussian action for the fluctuations about the saddle point, respectively.

\subsection{Comparison to experiments}
As we have argued in the main text, using the RPA result for the density fluctuations leads to an overdamped mode at imaginary frequency $\omega\sim i q^3 V(q)$ (corresponding to the relaxation of density fluctuations), which is subdiffusive for short-range interactions at small layer separation. For the mode with $\omega\sim i q^3$, this leads to a tunneling current $I\sim \tn{exp}[-(V_0/V)^{1/2}]$ at small currents and voltages, where $V_0$ is a constant. In Fig. \ref{expt} we show fits to the current, normalized with respect to the peak current $I_{\tn{max}}$ (which arises from different physics; as described in the main text and Ref.~\cite{JKJ17}) as a function of $1/V^{1/2}$ on a semi-logarithmic scale. It is clear that the fits capture the behavior well, which provides strong evidence for the applicability of the RPA results to the experimental regime of main interest in this paper.

\begin{figure}
\begin{center}
\includegraphics[width=0.8\columnwidth]{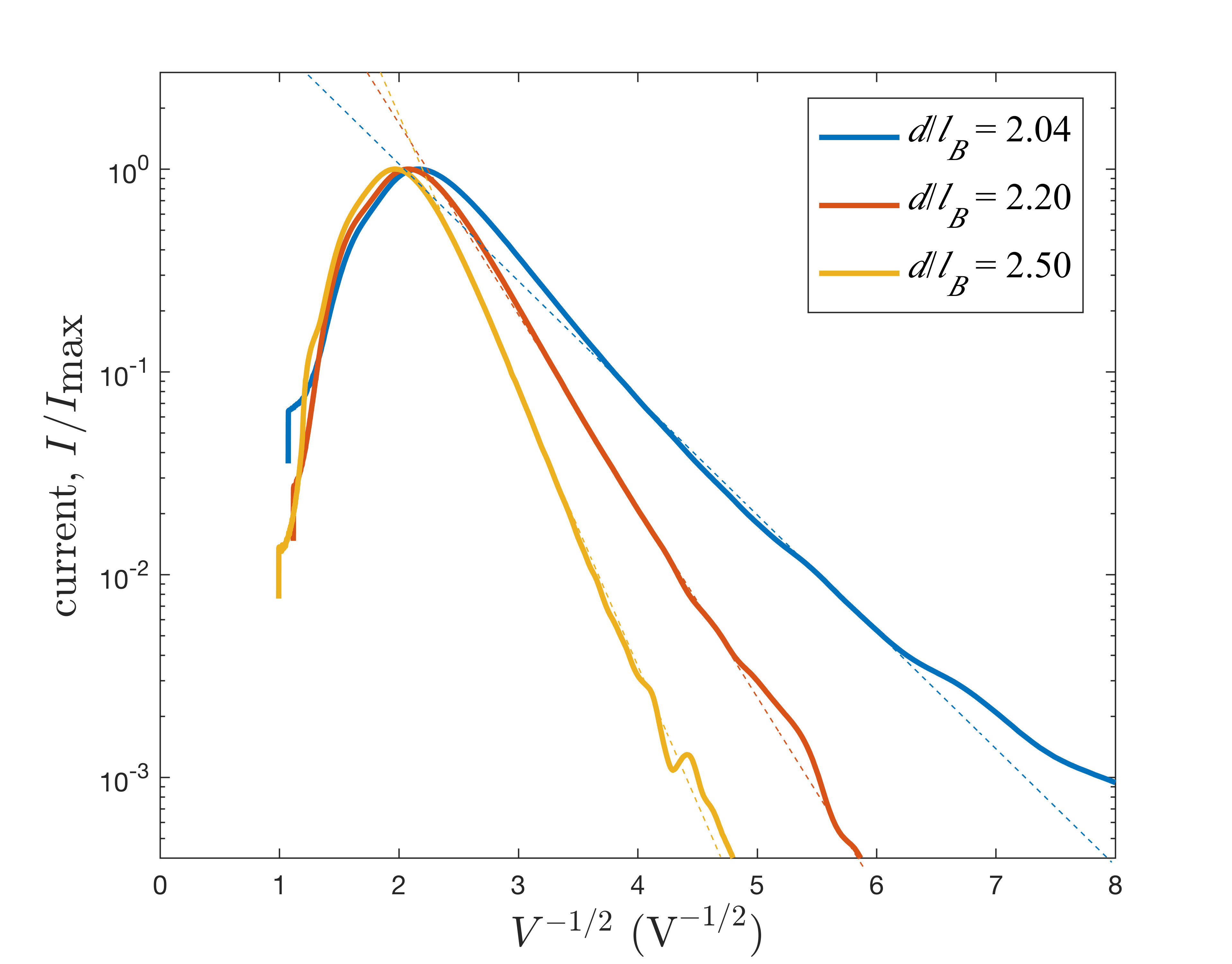}
\end{center}
\caption{A plot of the current, $I$, normalized by $I_{\tn{max}}$ (the peak current) as a function of $1/V^{1/2}$ for three different (small) values of $d/l_B$, where $d$ is the layer separation and $l_B = \sqrt{\hbar c/(e B_\perp)}$ is the magnetic length, at $B_{\Vert}=0$. (Data supplied by the Eisenstein group \cite{JPE18}; see also Ref.~\cite{JPE16}).}
\label{expt}
\end{figure} 

\end{widetext}

\end{document}